\newcommand{\be}{\begin{equation}}
\newcommand{\ee}{\end{equation}}
\newcommand{\bea}{\begin{eqnarray}}
\newcommand{\eea}{\end{eqnarray}}
\documentstyle[12pt]{article}
\begin{document}
\thispagestyle{empty}
\vspace{20cm}

\begin{center}

{\large\bf EXTENSIONS OF THE N=2 SUPERSYMMETRIC {~}{~~~~}
 a=-2 BOUSSINESQ HIERARCHY}
\vspace{1.5cm} \\
 Z. Popowicz \footnote{E-mail: ziemek@ift.uni.wroc.pl}
\vspace{1.0cm} \\
{Institute of Theoretical Physics, University of Wroclaw \\
pl.M.Borna 9, 50-204 Wroclaw, Poland}  \vspace{1.5cm}
\end{center}

\noindent{\bf Abstract.}
We present two different Lax operators for a manifestly N=2 
supersymmetric  extension of "a=-2" Boussinesq hierarchy .
The first is the supersymmetric generalization of the Lax operator of 
the 
Modified KdV equation. The second is the generalization of the 
supersymmetric Lax operator of the  $N=2$ supersymmetric a=-2 KdV 
system. 
The gauge transformation of the first Lax operator provide the
Miura link between the "small" N=4 supersymmetric conformal algebra 
and
the supersymmetric $W_{3}$ algebra .

\newpage

\section{Introduction}

The integrable hierarchies of differential equations in 1+1 
dimensions 
occupy an important place in diverse branches of theoretical physics 
as
exactly solvable models of fundamental physical phenomena ranging from
nonlinear hydrodynamics to string theory [1-3]. Recently N=2 
supersymmetric integrable hierarchies have attracted much attention  
(see [5-17] for example). This interest is motivated by both pure 
mathematical reasons and possible physical applications of these 
systems 
in  non-pertubative 2D supergravity, matrix models, etc. 

The integrable supersymmetric extensions of the KdV hierarchy bear an 
intimate relation to superconformal (super Virasoro) algebras via 
their 
second Hamiltonian structure [5]. This connection has been extended 
to more 
complicated algebra as  N=2 super $ W_{3}$ algebra. Due to it
three different super N=2 Boussinesq equations have been found [11-
12].
The Lax formulation of these systems has been given however only for 
two 
cases [12,13,14,9]. Recently Liu [15] constructed the N=1 super 
$W_{3}$ 
algebra via the second Gel'fand-Dickey bracket from the N=1 
supersymmetric Lax operator. Moreover he conjectured that this Lax 
operator producesome N=2 supersymmetric extension of the Boussinesq 
equation and the corresponding algebra coincides with the N=2 super 
$W_{3}$ algebra.

On the other side up to now, most efforts were focused on studing N=1
or N=2 supersymmetric systems. The higher supersymmetry N=4 have been 
used to supersymmetrization of the Liouville and Korteweg - de Vries 
system. Recently Delduc, Ivanov and Krivonos [7] has shown that the 
N=4 
super KdV equation could be written down in terms of the N=2 
superfields. 
Such system are called "quasi" N=4 Susy KdV hierarchies for which the 
Lax 
operator has been found also [8]. It appeared that the so called 
"small" 
supersymmetric conformal algebra $(SCA)$ is responsible for the  
the second hamiltonian structure of this hierarchy. Moreover this 
algebra is connected via the Miura transformation with the $N=2$ 
supersymmetric $W_{3}$ algebra [8]. 

In this letter we would like to study the relationship between the 
generalized Boussinesq and "quasi" N=4 Susy KdV hierarchy .
We show that similarly to the KdV hierarchy it is possible to 
construct new hierarchy for the  supersymmetric $a=-2$ Boussinesq 
equation. The investigations of such generalizations is interesting 
from the several reasons. One of them is the  possibility of the 
construction of a new generalization of the  constrained 
Kadomtsev-Petviashvili hierarchy [18]. Indeed, in the so called 
bosonic 
sector of our generalized supersymmetric Boussinesq hierarchy we 
recover 
new integrable hierarchy, different than this considered by Melnikov 
[19].

In order to end this we consider two different Lax operators. The 
first 
is a new and in the particular case coincides with the Lax operator 
considered by Liu [15]. The second is the generalization of the Lax
operator considered by Delduc and Gallot [8]. 

Moreover as the byproduct of our analysis we show that, in the 
particular 
case, our first operator give us new Lax representation of the 
"quasi" 
$N=4$ supersymmetric KdV system also. We show that the Lax operator 
of 
the supersymmetric generalization of the Boussinesq equation is gauge 
equivalent with the Lax  operator of the "quasi" $N=4$ supersymmetric 
KdV 
system. This gauge transformation define us the Miura transformation 
between "small" $SCA$ algebra and supersymmetric $W_{3}$ algebra.

Finally we show that the modification of the chirality conditions 
in the  Lax operator of the "quasi" $N=4$ Susy KdV hierarchy 
create new Lax opeartor which generate the supersymmetric Boussinesq 
hierarchy 

The hierarchies constructed by these 
operators constitue the hamiltonian structures and are connected by 
simple 
transformation among themselves. We explicitely constructed first 
Hamiltonian structure for these hierarchies.

Let us mention that the similar construction us our for 
the supersymmetric $N=2$ Boussinesq equation but for the $a=-1/2$ 
case 
has been carried out in [10]. However there is a basic difference 
between 
our aproach and this presented in [10] where authors included to the 
generalization, different from our, conformal dimensional chiral and 
anti-chiral superfields.

All computation presented in this paper has been carried 
out with the help of the symbolic computer language Reduce [20] and 
utilizing the package Susy2 [21].

\section{Notation}
The basic objects in the supersymmetric analysis are superfileds and 
the 
supersymmetric derivatives. The Taylor expansion of the superfield 
with 
respect to the $\theta$ is \be
\phi(x,\theta_{1},\theta_{2})=w(x)+\theta_{1}\xi_{1}+\theta_{2}\xi_{2}
\theta_{2}\theta_{2}u,
\ee
where the fields $w,u$ are to be interpreted as the boson (fermion) 
fields 
for the superboson (superfermion) field while $\xi_{1},\xi_{2}$, as 
the 
fermion (boson) for the superboson (superfermions) respectively. The
superderivatives are defined as
\bea
D_{1}=\partial_{\theta_{1}} -\frac{1}{2} \theta_{2}\partial,\\ \
D_{2}=\partial_{\theta_{2}}-\frac{1}{2} \theta_{1}\partial,
\eea
and satisfy $D_{1}^{2}=D_{2}^{2}=0$ and $ D_{1}D_{2} + 
D_{2}D_{1} =-\partial$.

Below we shall use the following notation: $(D_{i}F)$ denotes the 
outcome
of the action of the superderivative on the superfield F, while 
$D_{i} F$ 
denotes the action itself of the superderivative on the superfiled F.

We use, in the  next an obvious relation
\be
\int dx d\theta_{1} d\theta_{2} A = \int dx d\theta_{1} (D_{2}A) ,
\ee
valid for an arbitrary superfuction $A$ which rapidly vanishes when x 
$\rightarrow$$\pm\infty$. Notice that this identity can be 
interpreted as the supersymmetric analog of the Stokes formula [4].

\section{Supersymmetric N=2 Boussinesq equation}

This equation is written down as [12]
\newpage

\bea
\frac{d}{dt} V & = & 2T_{x} + a(2(([D_{1},D_{2}]V_{x}) + 4VV_{x}) \; 
, \ \\
\frac{d}{dt} T & = & -2V_{xxx}+([D_{1},D_{2}] 
T_{x})+10(D_{1}V)(D_{2}V)_{x} 
-2(V[D_{1},D_{2}]V)_{x} +4V^{2}V_{x} \cr
& &-2V_{x}([D_{1},D_{2}]V) + 4V^{2}V_{x} 
+(5-2a)((D_{1}V)(D_{2}T)+(D_{2}V)(D_{1}T)) +\cr
& & (8+4a)V_{x}T + (3+2a)VT_{x},
\eea
Here $a$ is an arbitrary parameter. As was shown in [11,12] for three
values of parameter $a=-2,1/2,-5/2$ this equation posseses at 
least five nontrival conserved currents. The Lax operator has been 
found only for two values of $a$ ($a= -1/2, -2$) [13,14,12,9] and 
hence, for these values, this system is integrable. 

In the next we consider the $a=-2$ case only. Then the equations (5-6)
can be reduced to much simpler form. Indeed if we shift the T 
superboson 
to the new one W \be 
T \rightarrow W+2([D_{1},D_{2}]V)+16V^{2},
\ee
and rescale in an appriopriate way the time, we obtain the following 
system 
of equations
\bea
\frac{\partial}{\partial t} V &=& 2W_{x}, \, \ \\
\frac{\partial}{\partial t} W &=& ([D_{1},D_{2}]W_{x}) + VW_{x}+
    (D_{2}W)(D_{1}V) + (D_{1}W)(D_{2}V) .
\eea
We present here two differnt Lax operator for this equation.

The first is 
\be
L= \partial^{2} + V\partial - (D_{2}V)D_{1} + W + D_{1}\partial^{-1}
(D_{2}W),
\ee
while the second is 
\be
L=[D_{1},D_{2}]\partial - D_{1}VD_{2} - D_{2}VD_{1} + 
\frac{1}{2}[D_{1},D_{2}]\partial^{-1} W 
+\frac{1}{2}W[D_{1},D_{2}]\partial^{-1}.\ee

The corresponding Lax pair which, produces equation (8-9) 
for both cases is
\be
\frac{d}{dt} L= [ L_{\geq 1},L],
\ee 
where ${\geq 1}$ denotes purely (super) differential part of the 
operator. We show, in the last chapter, the connection of the 
operator (11) 
with the Lax operator considered in [9].

As we see our first Lax operator (10), strictly speaking, is the $N=1$
supersymmetric. It posess the first Susy-derivative only.
From the point of view of $N=2$ supersymmetric theory 
it is degenerated susy-differential operator.
Assuming that $(D_{2}V)=H$ and $(D_{2}W)=T$ we recover the Lax 
operator
considered by Liu in [15].

The conserved currents, for the first Lax operator are defined by the 
standard (N=1) residue form as
\be
I_{n} = \int dx d\theta_{1} tr L^{\frac{n}{2}}
\ee
where $tr$ denotes the coefficient standing before $\partial^{-1} 
D_{1}$.
Now using the formula (4) we can cast these currents to the 
evident N=2 supersymmetric form.

Indeed. It is easy to noticed that an arbitrary power of our Lax 
operator
(10) commutes with the $D_{2}$ operator.
Let us present symbolicaly the expansion of 
\be 
L^{k} \rightarrow ... B + D_{1}\partial^{-1} F + ...
\ee
and
\be
[L^{k},D_{2}] \rightarrow ... (D_{2}B) +F + D_{1}\partial^{-
1}(D_{2}F) +
...=0,
\ee
for the zero and first pseud-Susy-derivative terms only . 
The superfunction $B$ and $F$ could be explicitely computed from the 
Lax 
operator (10). We conclude, from these expansions, that $F=-(D_{2}B)$ 
.
Therefore we can use the analog of Stokes formula (4) in order to 
bring 
the currents to the evidently $N=2$ supersymmetric form.

This supersymmetric generalization is bihamiltonian. The second 
hamiltonian structure is connected with the supersymmetric $W_{3}$ 
algebra 
[12] while first has simple form and reads as [13,12]
\vspace{0.5cm}

\be
P=\pmatrix{ 0, & 2\partial \cr 2\partial ,& [D_{1},D_{2}]\partial
+V\partial -(D_{1}V)D_{2} - (D_{2}V)D_{1})}.
\ee
\vspace{0.4cm}

\section{Susy N=2 Boussinesq hierarchy}
The first generalization which we would like to study  is connected 
with 
the Lax operator (10) and has the following form
\bea
L &=& \partial^{2} + V\partial -(D_{2}V)D_{1} + W 
+\partial^{-1}D_{1}(D_{2}W)-  \cr
& & \sum_{i=1}^{m}(F_{i}D_{1}\partial^{-1}(D_{2}G_{i})-
\partial^{-1}D_{1}(F_{i}(D_{2}G_{i})).
\eea
Here $m$ pairs of superbosons chiral and anit-chiral fields $F_{i}$ 
and 
$G_{i}$ satysfying
\be
(D_{1}G_{i})=(D_{2}F_{i})=0 .\ \\
\ee
with dimension $1$ were introduced.

We have checked that this Lax pair gives rise through the Lax 
pair (12) to the new hierarchy of the evolution equation.
When all $F{_k}=0$ and $G_{k}=0$, the Lax operator reduces to the 
(10).
Explicitly the first three flows are 
\bea
\frac{\partial}{\partial t_{1}} V = V_{x} ,{~~~} 
\frac{\partial}{\partial t_{1}} W = W_{x}, {~~~}
\frac{\partial}{\partial t_{1}} F_{i} = F_{ix},{~~~}
\frac{\partial}{\partial t_{1}} G_{i} = G_{ix},
\eea

\bea
\frac{\partial}{\partial t_{2}} V &=& 2W_{x} ,
\eea

\newpage

\bea
\frac{\partial}{\partial t_{2}} W &=& ([D_{1},D_{2}]W_{x}) + W_{x}V 
+(D_{1}W)(D_{2}V)+(D_{2}W)(D_{1}V) - \cr
& & 2\sum_{i=1}^{m}((D_{2}G_{i})(D_{1}F_{i}))_{x},\ \\
\frac{\partial}{\partial t_{2}} F_{i} &=&F_{ixx} + VF_{ix} -
(D_{2}V)(D_{1}F_{i}), \ \\
\frac{\partial}{\partial t_{2}} G_{i} &=& -G_{ixx} + VG_{ix} -
(D_{1}V)(D_{2}G_{i}).
\eea

For the third flow we scaled the time $ t \mapsto -\frac{1}{4}t$ and 
obtained
\bea
\frac{\partial}{\partial t_{3}} V &=& \partial(-
V_{xx}+3(D_{2}V)(D_{1}V)+
\frac{1}{2}V^{3} \cr
&& - 6([D_{1},D_{2}]W) - 6WV + 
 12\sum_{i=1}^{m}(D_{2}G_{i})(D_{1}F_{i})),\ \\
 \frac{\partial}{\partial t_{3}} &W& = -4W_{xxx}-
\frac{3}{2}W_{x}V^{2}  
-6W_{x}W \cr
&& -3\Big((D_{1}W)(D_{2}V)\Big)_{x} - 3(D_{1}W_{x})(D_{2}V) \cr
&& +3\Big((D_{2}W)(D_{1}V)\Big)_{x} + 3(D_{2}W_{x})(D_{1}V) \cr
&& -3\Big(([D_{1},D_{2}]W)V\Big)_{x} - 3([D_{1},D_{2}]W_{x})V \cr
&&-3V\Big((D_{1}W)(D_{2}V)+(D_{2}W)(D_{1}V)\Big) +\cr
&& 6\sum_{i=1}^{m}\Big( 2V\Big((D_{2}G_{i})(D_{1}F_{i})\Big)_{x} 
-2\Big(G_{ix}F_{ix}\Big)_{x} \cr
&&(D_{2}V)(D_{1}F_{i})G_{ix} -(D_{1}V)(D_{2}G_{i})F_{ix}\Big), \ \\
\frac{\partial}{\partial t_{3}} F_{i} &=& -4F_{ixxx} - 6WF_{ix} 
-3\Big(VF_{ix}\Big)_{x}  -3VF_{ixx} \cr
&&+ 3\Big((D_{2}V)(D_{1}F_{i})\Big)_{x} + 
3(D_{2}V)(D_{1}F_{ix})+6(D_{2}W)(D_{1}F_{i})\cr
&& +3(D_{2}V)(D_{1}F_{1})V -\frac{3}{2}V^{2}F_{ix}, \ \\
\frac{\partial}{\partial t_{3}} G_{i} &=& -4G_{ixxx} -6WG_{ix} 
+3\Big(VG_{ix}\Big)_{x} + 3VG_{ixx} \cr
&&-3\Big((D_{1}V)(D_{2}G_{i})\Big)_{x} -3(D_{1}V)(D_{2}G_{ix}) +
6((D_{1}W)(D_{2}G_{i}) \cr
&& +3(D_{1}V)(D_{2}G_{i})V -\frac{3}{2}V^{2}G_{ix} .
\eea

If we transform our $W$ superfield to the following form 
\be
W \mapsto \sum_{i=1}^{m} F_{i}G_{i},
\ee
then we obtain that the transformed Lax operator 
\newpage

\be 
L=\partial^{2} +V\partial -(D_{2}V)D_{1}+\sum_{i=1}^{m}
\Big( F_{i}G_{i} +F_{i}\partial^{-1}D_{1}(D_{2}G_{i}) \Big),
\ee
generates for $m=1$ the so called $"quasi"$ N=4 Susy KdV system 
considerd
in [8] . In that manner we obtained different Lax representation, 
than this 
considered in [8] for these equations.

Now let us assume that $m=1$ and eliminate the  field $F_{1}$ 
in the operator (29). In order to do it we assume that $F_{1} \neq 0$ 
and gauge this operator to the new one 
\be 
\overline L \Rightarrow \frac{1}{F_{1}} L F_{1}.
\ee
We see that this transformed Lax operator has the same structure 
as the operator (10) if we make the following identification
\be
\overline V = V + 2\frac{F_{1x}}{F_{1}} ,
\ee
\be
\overline W = F_{1}G_{1} + \frac{F_{1xx}}{F_{1}} 
+\frac{VF_{1x}}{F_{1}}-
\frac{(D_{2}V)(D_{1}F_{1})}{F_{1}}.
\ee
These formulas define us the Miura transformation between "small" 
$N=4$ 
supersymmetric conformal algebra and supersymmetric $W_{3}$ algebra.
This transformation coincides with this considerd in  [8].

We succeed find the first Hamiltonian structure for our equations 
(20-23).It has the following form
\be
PP=\pmatrix{ P & 0 \cr 0 & II},
\ee
where $P$ is defined by (16) while $II$ is $2m$ dimensional matrix
\be
II=\pmatrix{ 0 & I \cr -I & 0},
\ee
and $I$ is an identity m-dimensional matrix.

This Poisson tensor produces a new hierarchy of integrable equations 
of the form
\be
\frac{\partial}{\partial t_{n}} (V,W,F_{1},..G_{1},,)^{tp}=
PP grad(I_{n}),
\ee
where $tp$ denotes transposition, $I_{n}$ is defined by (13) and 
$grad$ 
denotes the functional gradient. Explicitly for our  
equations (20-23) it has the following form
\be
I_{2}=\frac{1}{2}(W^2+2\sum_{k=1}^{m}(G_{kx}F_{kx}-VG_{k}F_{kx}+
G_{k}(D_{2}V)(D_{1}F_{k}))).\
\ee

It is interesting to check what kind of the hierarchy we 
obtain if we add chiral and antichiral fields to the Lax operator 
(11), in 
a simillar manner as for Lax operator (10). We have 
checked several possibilities and concluded that this generalization 
can be 
cast into the form
\bea
L &=& [D_{1},D_{2}]\partial + D_{1}VD_{2} + D_{2}VD_{1}+
[D_{1},D_{2}]\partial^{-1} W +W[D_{1},D_{2}]\partial^{-1}+ \cr
&& + \sum_{i=1}^{m}\Big( F_{i}\partial^{-1}[D_{1},D_{2}] G_{i}+
G_{i}\partial^{-1}[D_{1},D_{2}]F_{i}\Big).
\eea
If we assume the chirality conditions on the superfields $F_{i}$ and 
$G_{i}$ (eq.18) then we obtain the following second flow:
\bea
\frac{\partial}{\partial t}V &=& 2W_{x} +
2\sum_{i=2}^{m}(F_{i}G_{i})_{x},\; \ \\
\frac{\partial}{\partial t}W &=& -([D_{1},D_{2}]W_{x})+W_{x}V+
(D_{1}W)(D_{2}V) + (D_{2}W)(D_{1}V),\; \ \\
\frac{\partial}{\partial t}F_{i} &=& F_{ixx} + VF_{ix} 
-(D_{1}V)(D_{2}F_{i}), \;\ \\
\frac{\partial}{\partial t} G_{i} &=& -G_{ixx} +VG_{ix} 
-(D_{2}V)(D_{1}G_{i}).
\eea

Interestingly these equations are  equivalent with the 
equations (20-23) if we transform the superboson $W$
\be
W \rightarrow W + \sum_{k=1}^{m} G_{k}F_{k}.
\ee

Hence we can state that this system is also the hamiltonian system.

\section{From SUSY KdV Hierarchy to the Boussinesq Hierarchy}

We have seen in the previous section that  it was possible to 
describe 
two different hierarchies, "quasi" $N=4$ Susy Kdv and Boussinesq, 
using 
one Lax operator only. We now show that our generalized Boussinesq 
hierarchy can be described also by the Lax operator of the "quasi" 
N=4 
Susy KdV with much weaker condition than the chirality assumption 
(18).

In order to do it, let us first consider the Lax operator (29) where 
we 
do not assume any chirality conditions. Then in order to obtain the 
self-consistent equation of motion, it appeares that we can assume 
much 
weaker conditions than (18). Indeed, it is enough to assume that 
\be
(D_{2}F_{i})=0, {~~~}{~~}{~~~} 
F_{ix}\Big(G_{ix}+(D_{1}D_{2}G_{i})\Big)=0,
\ee
for all $i$.

We can find two different solutions for the last equations. 
The first one is 
\be
F_{1} = const, {~} {~} {~} G_{1}=W,
\ee
while for $i=2,3,..m$ we assume the conditions (18).
The second solution is given by the conditions (18) for all $i$. 
In the first case we obtain the generalization of the Boussinesq 
hierarchy,
while for the second case we got new Lax operator for the "quasi" N=4 
Susy KdV system.

Let us now consider our second Lax operator which we rewrite it as 
follow
\bea
L &=& [D_{1},D_{2}]\partial + D_{1}VD_{2} + D_{2}VD_{1}+\cr
&& + \sum_{i=1}^{m}\Big( F_{i}\partial^{-1}[D_{1},D_{2}] G_{i}+
G_{i}\partial^{-1}[D_{1},D_{2}]F_{i}\Big).
\eea
We do not assume the (anti)chirality conditions (18) on the fileds
$F_{i}$ 
and $G_{i}$.This Lax operator generates the self-consistent equations 
provided\be
(D_{2}F_{i})(D_{1}D_{2}G_{i}) + (D_{2}G_{i})(D_{1}D_{2}F_{i}) =0,
\ee
\be
 \Big( (D_{1}F_{i})((D_{1}D_{2}G_{i}) +G_{ix})+
(D_{1}G_{i})((D_{1}D_{2}F_{i})+F_{ix})\Big)=0,
\ee
for all $i$.

We obtained  weaker conditions than (18) also. Moreover, the 
constraints 
(46-47) have the same solution as in the previous case.
Therefore, similarly to the previous case we conclude  that our first 
solution give us the generalization of the Boussinesq equation while 
the 
second solution the "quasi" N=4 Susy KdV Lax operator.

Finally let us present the connection of our Lax operator (11) with 
the Lax
operator of the Boussinesq equation considered in [9]. First let us 
notice that the Susy N=2 $a=-2$ KdV posseses four different Lax 
operators
\bea 
L_{1} &=& \partial^{2} + D_{1}VD_{2} - D_{2}VD_{1}, \ \\
L_{2} &=& D_{1}\Big(\partial +2V\Big)D_{2},\  \\
L_{3} &=& \partial^{2} + D_{1}VD_{2} ,\ \\
L_{4} &=& \partial^{2} + V\partial -(D_{2}V)D_{1}.
\eea

These Lax operators generate the same Susy $N=2$ $a=-2$ KdV equation
\be
\frac{\partial}{\partial t} V =\partial \Big( -
V_{xx}+6(D_{2}V)(D_{1}V)
+2V^{3}\Big). \ \\
\ee

The first two Lax operators have been considered in [13,16] and they 
are
connected each other as
\be
L_{1} \mapsto L_{2} + L_{2}^{*},
\ee
where $*$ denotes the hermitean conjugation.

The last one is simple the supersymmetric version of the Lax operator 
of the 
Modified KdV equation and its generalization has been considered in 
the 
previous sections.

We succeed to generalize the second Lax operator and obtained Lax 
operator
\be
L=D_{1}(\partial +2V)D_{2} 
+\sum_{i=1}^{m}G_{i}\partial^{-1}D_{1}D_{2}F_{i} ,
\ee
which have been considered in [8] but now we do not assume any 
chirality .
However it appeare that we have to asume such in order to obtain the 
self-
consistent equations. In order to weak this assumption let us now 
consider 
a new Lax operator $LL$ constructed as
\be
LL=L-L^{*}.
\ee
This form has predicted the unexpected behaviour of the 
supersymetrical 
solitonic equations. We encounter several new phenomena, compare to 
the
classical case, during the supersymetrization. One of them is the 
observation that sometimes the linear combination of Lax operator and 
 
its hermitean conjugation gives us the equation which is different 
than 
this produced by Lax operator only. For example, the Lax operator [17]
\be
L=\partial +\partial^{-1}[D_{1},D_{2}] V,
\ee
produces the Susy $N=2$ $a=1$ KdV system while its linear combination 
as 
(54) gives the Susy $N=2$ $a=4$ KdV system.

The operator (55) where $L$ is defined by (54) coincides with the Lax 
operator (11). Indeed we can repeat the previous arguments to this 
Lax 
operator to establish the same connection with the Boussinesq 
hierarchy 
also.

On the other hand it is also possible to decompose this Lax operator 
in a 
different manner. Namely, we introduce the operator 
\be
N=D_{1}\Big( \partial + V + D_{2}\partial^{-1}W\partial^{-1}D_{1} +
\sum_{i=1}^{m} G_{i}\partial^{-1}F_{i}\Big)D_{2},
\ee
where we now assume the chirality conditions (16) on the fields 
$F_{i}$ and
$G_{i}$ .
We quickly realize that it is possible to write $LL$  as $LL=N-N^{*}$.
When all $F_{i}=0$ and $G_{i}=0$ than the operator $N$ is exactly the 
Lax operator introduced in [9].

\end{document}